\documentclass[12pt,preprint]{aastex}
\usepackage{natbib}
\shorttitle{Collapse of Filamentary Molecular Clouds}
\shortauthors{Tilley \& Pudritz}
\begin{document}
\title{Gravitational Collapse of Filamentary Magnetized Molecular Clouds}
\author{David A. Tilley\altaffilmark{1} and Ralph E. Pudritz\altaffilmark{2}}
\affil{Department of Physics and Astronomy}
\affil{McMaster University, Hamilton, ON L8S 4M1}
\altaffiltext{1}{tilley@physics.mcmaster.ca}
\altaffiltext{2}{pudritz@physics.mcmaster.ca}
\begin{abstract}
  We develop models for the self-similar collapse of magnetized isothermal cylinders.  We find solutions for the case of a fluid with a constant toroidal flux-to-mass ratio ($\Gamma_{\phi}=\mathit{constant}$) and the case of a fluid with a constant gas to magnetic pressure ratio ($\beta=\mathit{constant}$).  In both cases, we find that a low magnetization results in density profiles that behave as $\rho\propto r^{-4}$ at large radii, and at high magnetization we find density profiles that behave as $\rho\propto r^{-2}$.  This density behaviour is the same as for hydrostatic filamentary structures, suggesting that density measurements alone cannot distinguish between hydrostatic and collapsing filaments---velocity measurements are required.  Our solutions show that the self-similar radial velocity behaves as $v_r\propto r$ during the collapse phase, and that unlike collapsing self-similar spheres, there is no subsequent accretion (i.e. expansion-wave) phase.  We also examine the fragmentation properties of these cylinders, and find that in both cases, the presence of a toroidal field acts to strengthen the cylinder against fragmentation.  Finally, the collapse time scales in our models are shorter than the fragmentation time scales.  Thus, we anticipate that highly collapsed filaments can form before they are broken into pieces by gravitational fragmentation.
\end{abstract}
\keywords{ISM: clouds --- ISM: kinematics and dynamics --- ISM: magnetic fields --- MHD --- stars: formation}

\section{INTRODUCTION}
  	Filamentary structures are commonly seen in giant molecular clouds as sites in which are embedded star-forming clumps.  
These structures are threaded by magnetic fields \citep{heiles93,matthews00,falgarone01}, which can support the filaments against gravitational collapse.
While the mechanism by which filaments form is not well understood, it has been suggested that these filamentary structures occur as the result of the fragmentation of the parent molecular cloud through turbulent motions \citep{klessen00a,ostriker01}.  Recent theoretical work \citep{fiege00a,fiege00b} suggests that helical magnetic fields may play an important role in determining the structure of a filament.

	The study of the collapse of a spherical pre-stellar core has been greatly benefitted by the study of one-dimensional models \citep{larson69,penston69,shu77,foster93},  
which have become very useful in identifying the physical origin of a particular structural phenomenon.  By leaving out most of the complicated and analytically intractable physics which may occur in the full three-dimensional system, these models allow us to identify the physical processes which lead to the observed structure, and can be compared to both numerical simulations and observational data of molecular cloud cores.  The spherical self-similar models feature an initial gravitational collapse phase that ends with the appearance of a singularity in the density field.  This is followed by an accretion phase, wherein an expansion wave propagates outward from the newly formed core enabling the infall of the remaining envelope.
Similar analyses of the process of filament formation and evolution have not been as well studied, despite the frequency of filamentary structures forming in numerical simulations \citep{porter94,klessen00a,ostriker01}, and the abundance of filaments in observations of molecular clouds (e.g. \citeauthor{falgarone01} \citeyear{falgarone01}).

	\citet{stodolkiewicz63} found solutions for the equilibrium structure of an isothermal cylinder, for a fluid where the ratio of gas pressure to magnetic pressure is constant.  Changing the value of this ratio has the effect of changing the scale radius of the solution, which falls off as $(r/r_0)^{-4}$.  \citet{ostriker64} found identical solutions for the unmagnetized isothermal cylinder.

	\citet{miyama87} derived a set of self-similar solutions for an unmagnetized isothermal cylinder.  These solutions have an infall velocity which is proportional to the distance from the axis, and a density structure that has the same form as the unmagnetized equilibrium filaments of \citet{stodolkiewicz63} and \citet{ostriker64}, but with a scale radius that decreases with time and a central density that increases with time.

	Recently, \citet{hennebelle02} explored a set of self-similar solutions for a rotating, magnetized filament which may undergo collapse in the axial direction, in addition to radial collapse.  This set of solutions include the equations which we will derive in the next section, although \citet{hennebelle02} does not explore the case where there are only purely radial motions.  This is the focus of our analysis.

	The goal of this paper is to extend the calculations of the self-similar collapse of an isothermal cylinder to include magnetic and rotational effects.  
We show that one has to choose a relationship between the magnetic field strength and the density.  When this is done, we find that the density and magnetic field structures of collapsing cylinders is identical to that of the corresponding hydrostatic structure, with only the scale radius and normalization of the quantities changing in time.  The important consequence of this for observations is that only careful large-scale velocity measurements can discriminate between static and evolving filamentary structure, unlike the situation for accreting spheres.  We derive and solve our set of self-similar equations in Section \ref{sectioneq}. These solutions will be discussed in Section \ref{sectionresults}.

\section{THE SELF-SIMILAR COLLAPSE EQUATIONS}\label{sectioneq}

	We begin our analysis with the equations of ideal magnetohydrodynamics for a self-gravitating gas:
\begin{eqnarray}
\frac{\partial \rho}{\partial t} + \mathbf{\nabla\cdot}\left(\rho\mathbf{v}\right) & = & 0 \label{continuity}\\
\frac{\partial \mathbf{B}}{\partial t} & = & \mathbf{\nabla \times} \left(\mathbf{v \times B}\right) \\
\rho \frac{\partial \mathbf{v}}{\partial t} + \rho\left(\mathbf{v\cdot\nabla}\right)\mathbf{v} & = & -\mathbf{\nabla} P + \frac{1}{\mu}\left[\left(\mathbf{B\cdot\nabla}\right)\mathbf{B}-\frac{1}{2}\mathbf{\nabla}\left(\mathbf{B}^2\right)\right]-\rho\mathbf{\nabla}\Phi \\
\mathbf{\nabla}^2\Phi & = & 4 \pi G \rho \label{poisson}
\end{eqnarray}
In these equations, $\rho$ is the density, $\mathbf{v}$ is the velocity, $\mathbf{B}$ is the magnetic field, $P$ is the thermal pressure, for which we adopt the isothermal equation of state $P=c_s^2\rho$ (with a constant speed of sound $c_s$), and $\Phi$ is the gravitational potential.  

	The geometry of our filaments leads us to use a cylindrical coordinate system to solve the equations.  Our cylindrical symmetry causes all $\partial/\partial\phi$ and $\partial/\partial z$ derivatives to be zero, so that the resulting fields are functions of $\left(r,t\right)$ only.  This geometry allows us to replace the gravitational potential by a mass per unit length $m$, defined by
\begin{eqnarray}
\frac{\partial m}{\partial r} & = & 2\pi r\rho \label{rtmass}
\end{eqnarray}

  We can make the following substitutions, to change our equations from functions of ($r,t$) to separable functions of ($s,t$), for the self-similar variable $s=r/(c_s t)$.  These solutions will represent collapsing motions for $s<0$ (and hence $t<0$), and expanding motions for $s>0$.

\begin{eqnarray}
\rho(r,t) & = & \frac{1}{\pi G t^2}\tilde{\rho}(s)\label{rhortos}\\
m(r,t) & = & \frac{2 c_s^2}{G}\tilde{m}(s)\\
\mathbf{v}(r,t) & = & c_s \mathbf{\tilde{v}}(s)\\
\mathbf{B}(r,t) & = & \sqrt{\frac{\mu}{\pi G}}\frac{c_s}{t}\mathbf{\tilde{B}}(s)\label{brtos}
\end{eqnarray}
where $\tilde{\rho}(s)$ etc. are dimensionless functions of the self-similar variable $s$ alone.  Substituting these into Equations (\ref{continuity})-(\ref{rtmass}) (see Appendix \ref{appendixa} for details), 
\begin{eqnarray}
\tilde{v}_r & = & s\label{scont}\\
\frac{d\tilde{m}}{ds} & = & s\tilde{\rho}\label{smass}\\
\frac{d}{ds}\left(s\tilde{B}_r\right) & = & 0\label{sindr}\\
\frac{d}{ds}\left(\tilde{v}_{\phi} \tilde{B}_r\right) & = & 0\\
\tilde{B}_{r}\frac{d\tilde{B}_z}{ds} & = & 0 \label{smomz}\\
\tilde{B}_z & = & \frac{1}{s}\frac{d}{ds}\left(s\tilde{v}_z\tilde{B}_r\right)\label{sindz}\\
\frac{\tilde{\rho}\tilde{v}_{\phi}^2}{s} & = & \frac{d\tilde{\rho}}{ds}+\frac{4\tilde{m}\tilde{\rho}}{s}+\frac{\tilde{B}_{\phi}}{s}\frac{d}{ds}\left(s\tilde{B}_{\phi}\right)+\tilde{B}_z\frac{d\tilde{B}_z}{ds}\label{smomr}\\
\tilde{\rho}\tilde{v}_{\phi} & = & \frac{\tilde{B}_r}{s}\frac{d}{ds}\left(s\tilde{B}_{\phi}\right)\label{smomphi}
\end{eqnarray}

Before we proceed to solve Equations (\ref{scont})--(\ref{smomphi}), we will first recap the solutions to these equations in the case of $\mathbf{B}=0$.  Equation (\ref{smass}) remains unchanged, Equation (\ref{smomphi}) shows that $\tilde{v}_{\phi}=0$, and Equation (\ref{smomr}) reduces to $\mathrm{d}(\ln\tilde{\rho})/\mathrm{d}(\ln s)+4\tilde{m}=0$.  The solution to this, combined with Equation (\ref{smass}), is $\tilde{\rho} = 2 \zeta \left(1+\zeta s^2\right)^{-2}$ and $\tilde{m} = \zeta s^2 \left(1+\zeta s^2\right)^{-2}$, where $\zeta$ is a constant of integration.  This solution was first found by \citet{miyama87}.  It has two features worth highlighting:

\begin{enumerate}
  \item It has the same radial density structure as the equilibrium solution of \citet{stodolkiewicz63} and \citet{ostriker64}.
  \item The total enclosed mass per unit length (i.e. as $|s|\rightarrow\infty$) is the critical mass per unit length.
\end{enumerate}

	The asymptotic density behaviour of this solution tends to $s^{-4}$ at large values of s; we will show later that magnetic solutions tend to have this behaviour only in the limit of low magnetization.

	There are two classes of solutions to Equations (\ref{scont})--(\ref{smomphi}) when the effects of magnetic fields are included, which we can separate through the use of Equation (\ref{smomz}):
\begin{enumerate}
\item $\tilde{B}_r = 0$: In this case, we can show from Equation (\ref{sindz}) that $\tilde{B}_z =0$ and from Equation (\ref{smomphi}) that $\tilde{v}_{\phi}=0$ as well.  This solution also has the consequence that our system of equations is underspecified, as $\tilde{B}_{\phi}$ only appears in the radial momentum equation.  We will discuss the solutions to this below.
\item $\mathrm{d}\tilde{B}_z/\mathrm{d}s = 0$: If our $\tilde{B}_z$ field is constant, then from Equation (\ref{sindr}) we can see that we have a radial field $\tilde{B}_r=\tilde{B}_{r0}/s$ which is singular along the cylinder axis.
\end{enumerate}
	The latter solution is difficult to justify physically; even though the field is still divergence-free, the field lines are not closed because of the linear singularity at the axis of symmetry.  Henceforth we discard this possibility and focus our attention on Case 1.

	We now proceed to solve the remaining set of equations.  With the condition that $\tilde{B}_r=0$, we are left with
\begin{eqnarray}
\tilde{v}_r & = & s\\
\frac{d\tilde{m}}{ds} & = & s\tilde{\rho}\label{sfmass}\\
0 & = & \frac{d\tilde{\rho}}{ds} + \frac{4\tilde{m}\tilde{\rho}}{s} + \frac{\tilde{B}_{\phi}}{s}\frac{d}{ds}\left(s\tilde{B}_{\phi}\right)\label{sfmom}
\end{eqnarray}

We can integrate Equation (\ref{sfmom}) to obtain the following relationship between $\tilde{\rho}$, $\tilde{m}$ and $\tilde{B}_{\phi}$:
\begin{eqnarray}
\alpha & = & s^2\left(2\tilde{\rho}+\tilde{B}_{\phi}^2\right)+4\tilde{m}\left(\tilde{m}-1\right)\label{c1moma}
\end{eqnarray}
We can constrain the value of the constant $\alpha$ by examining the behaviour of the system on the axis.  Since $\lim\limits_{s\rightarrow0}\tilde{m}=0$, it follows that in the limit $s\rightarrow0$, $\lim\limits_{s\rightarrow0}s^2(2\tilde{\rho}+\tilde{B}_{\phi}^2)=\alpha$.

	The first term in this result can be evaluated by using Equation (\ref{sfmass}).  We can express the density as $\tilde{\rho} = A |s|^n$ as $s\rightarrow0$, as for sufficiently small $s$, one term in the expansion of $\tilde{\rho}$ will dominate the other terms.  From Equation (\ref{sfmass}), we can write the mass per unit length for this density behaviour as $\tilde{m}=A\int_0^{|s|} |s|^{n+1}\mathrm{d}|s|$.  For $n\le -2$, $\tilde{m}$ will not be finite in the limit $s\rightarrow0$, so with the constraint $n>2$, the first term in Equation (\ref{c1moma}) has the limit $\lim\limits_{s\rightarrow0} \tilde{\rho}s^2 = 0$.

\begin{figure}
\plotone{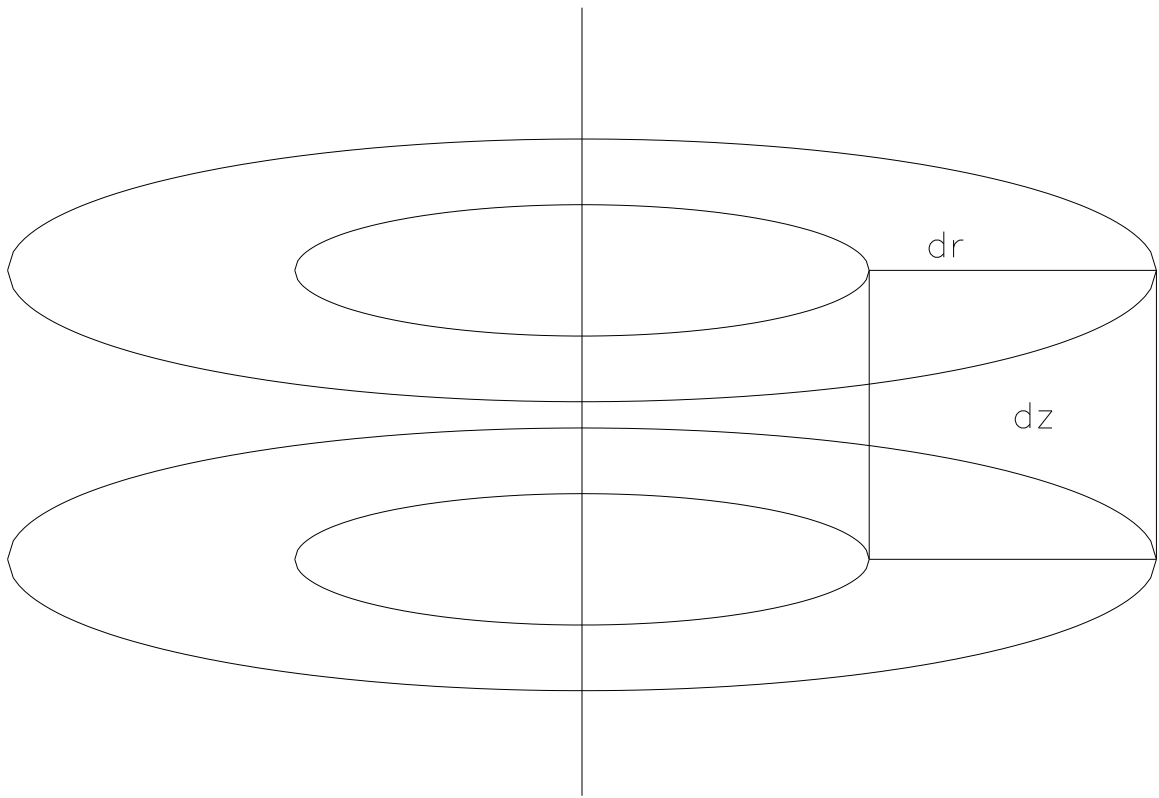}
\caption{Area element for the calculation of the toroidal magnetic flux.\label{magfluxschem}}
\end{figure}

	We can constrain the magnetic field behaviour in a similar manner, by examining the toroidal magnetic flux, $\Psi$.  We use the formulation of the magnetic flux of \citet{fiege00a}, such that $\Psi = \int \tilde{B}_{\phi} \mathrm{da}$, where $\mathrm{da=dr\;dz}$ is the area element through which the magnetic field passes (see Figure \ref{magfluxschem}).  If we consider the flux per unit length $\psi = \mathrm{d}\Psi/\mathrm{dz}$, and write the magnetic field as $\tilde{B}_{\phi} = C |s|^{n'}$ as $s\rightarrow0$, we find that $\psi = C\int_0^s |s|^{n'} \mathrm{d}s$.  For $n' \le -1$, $\psi$ is not finite, so we restrict $n' > -1$.  We thus establish that $\alpha = 0$ in Equation (\ref{c1moma}).

	The properties of the system can be examined if we solve for the mass per unit length in Equation (\ref{c1moma}).
\begin{eqnarray}
\tilde{m}_{\pm} & = & \frac{1}{2} \pm \frac{1}{2}\sqrt{1-s^2\left(2\tilde{\rho}+\tilde{B}_{\phi}^2\right)}\label{mpm}
\end{eqnarray}
We can identify two branches to this solution.  The minus branch $\tilde{m}_- \le 1/2$ corresponds to the minus sign in Equation (\ref{mpm}); the plus solution $\tilde{m}_+ \ge 1/2$ is the other branch.  The two branches meet at the half-mass point given by $s_{1/2}^2\left[2\tilde{\rho}(s_{1/2})+\tilde{B}_{\phi}^2(s_{1/2})\right]=1$.  Since $\tilde{m}(0)=0$, we know that the critical quantity $q_{\mathrm{crit}}=s^2(2\tilde{\rho}+\tilde{B}_{\phi}^2)\rightarrow 0$ as $s\rightarrow 0$.  Thus, we can determine that for every solution, the quantity $q_{\mathrm{crit}}$ must increase from $0\rightarrow1$, then begin to decrease again.  Thus, the minus branch begins at $\tilde{m}_-(0)=0$, then increases to $1/2$, then decreases again; the plus branch starts at $\tilde{m}_+(0)=1$, decreases to $1/2$, then increase again.

	Since the plus branch is always greater than $1/2$, it cannot represent the solution at $s=0$; similarly, at values of $|s|>|s_{1/2}|$, the mass per unit length must continue to increase (as a decreasing mass per unit length implies a negative density according to Equation (\ref{sfmass}), which is unphysical), and so the minus solution is unphysical in this regime.  Thus the physical solution must begin on the minus branch at low $s$, then at the half-mass point switch to the plus branch.  The maximum value of $\tilde{m}$ must occur when $\tilde{\rho}=0$.  If this occurs at a finite radius $s_{\mathrm{max}}$, then the total mass per unit length will be
\begin{eqnarray}
\tilde{m}_\mathrm{max} & = & \frac{1}{2}\pm\frac{1}{2}\sqrt{1-s_\mathrm{max}^2\tilde{B}_{\phi}^2(s_\mathrm{max})}
\end{eqnarray}
where the plus or minus sign is determined by whether $|s_\mathrm{max}| > |s_{1/2}|$ or $|s_\mathrm{max}| < |s_{1/2}|$, respectively.  	This $\tilde{m}_{\mathrm{max}}$ is identical to the reduced critical mass per unit length $m_{\mathrm{mag}}$ found by \citet{fiege00a} from the virial theorem, for a filament with a toroidal magnetic field.  

	Because of the indeterminacy in this solution as noted above, we need to specify a form for $\tilde{B}_{\phi}$ in order to make further progress.  We will perform this calculation for two plausible magnetic field configurations which commonly appear in star formation literature: one where the ratio of the thermal to magnetic pressure $\beta = 2\mu c_s^2 \rho / B_{\phi}^2 = 2\tilde{\rho}/\tilde{B}_{\phi}^2$ is constant, and one where the toroidal flux-to-mass ratio $\Gamma_{\phi} = \tilde{B}_{\phi}/(|s|\tilde{\rho})$ is constant \citep{fiege00a}.  We will discuss these two cases in turn.

\section{RESULTS}\label{sectionresults}

  There are two models for the magnetization of the gas in molecular clouds which are used extensively in the literature: one in which the magnetic pressure is proportional to the gas pressure, and one in which the magnetic flux through a fluid element is proportional to the mass of that element.  As we have the freedom in our set of equations to specify one of our variables, we choose to constrain the magnetic field through one of these relations.

	As we show in the Appendix, the structure of the self-similar collapse equations (\ref{sfmass})-(\ref{sfmom}) is identical in form to the equations for the hydrostatic solutions; the density and magnetic fields have identical radial structures.  The only difference between the static and collapse solutions is that the collapse solutions are time-dependent, and have an infall velocity.  It is remarkable that the radial velocity terms in the momentum, induction and continuity equations exactly cancels out the time-dependent terms.
A very different behaviour is seen in spherical symmetry, for which the collapse solutions have a different density structure than the static solutions.  This underlies the importance that the geometry has in determining the spatial structure of the collapsing filament.

	In self-similar cylindrical coordinates, the continuity equation (\ref{continuity}) restricts the radial velocity to a single solution, $v_r = r/t$.  Thus, this analysis can only represent a collapsing solution for $t<0$.  Furthermore, this implies that there is no equivalent to the accretion solution found in the self-similar collapse of a spherical isothermal fluid by \citet{larson69}, \citet{penston69}, and \citet{shu77}.  The collapse of filaments, therefore, is very different than the collapse of spheres.

\subsection{Case 1A: Constant $\beta$}\label{subsectionconstantbeta}
\begin{figure}
\plottwo{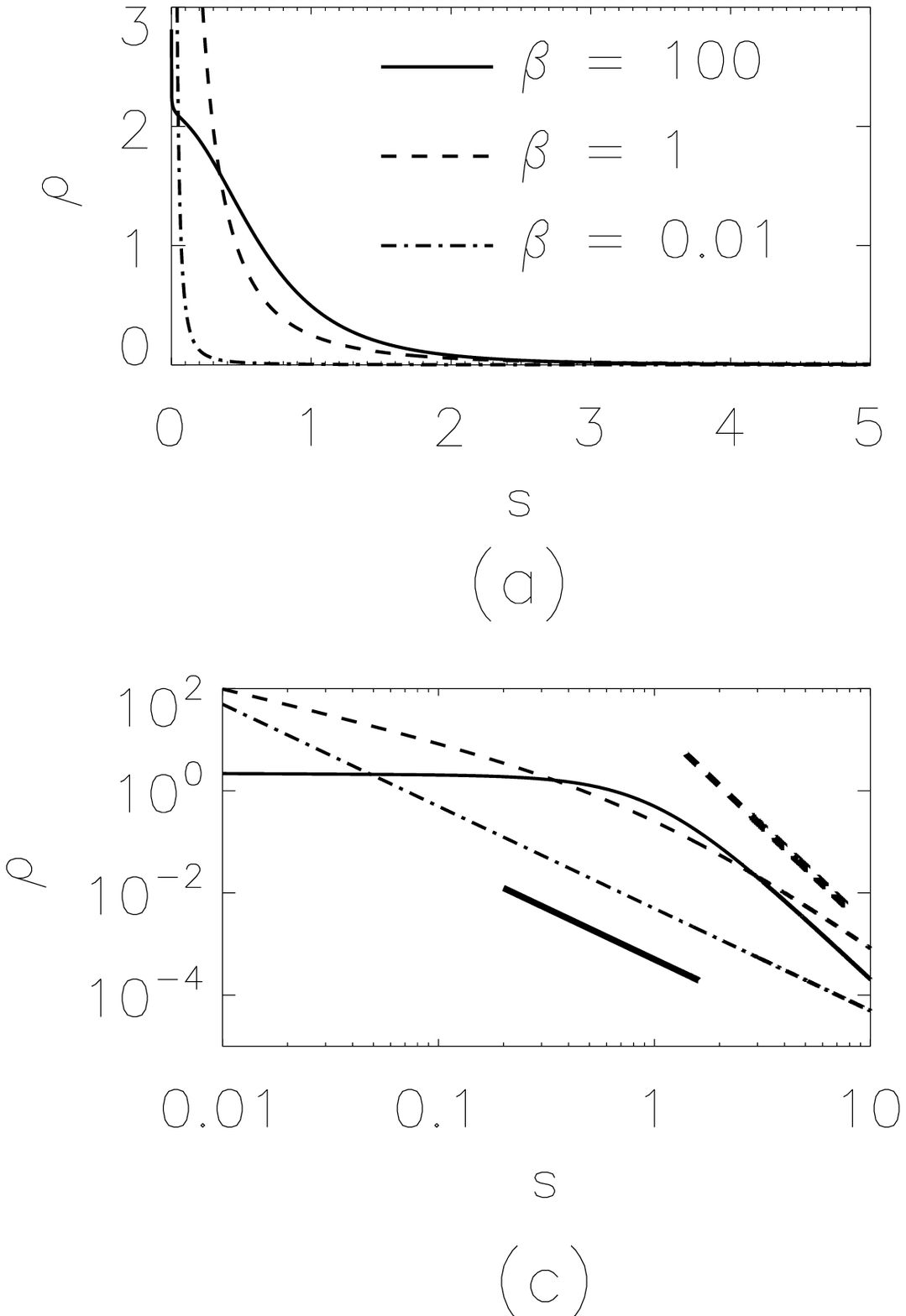}{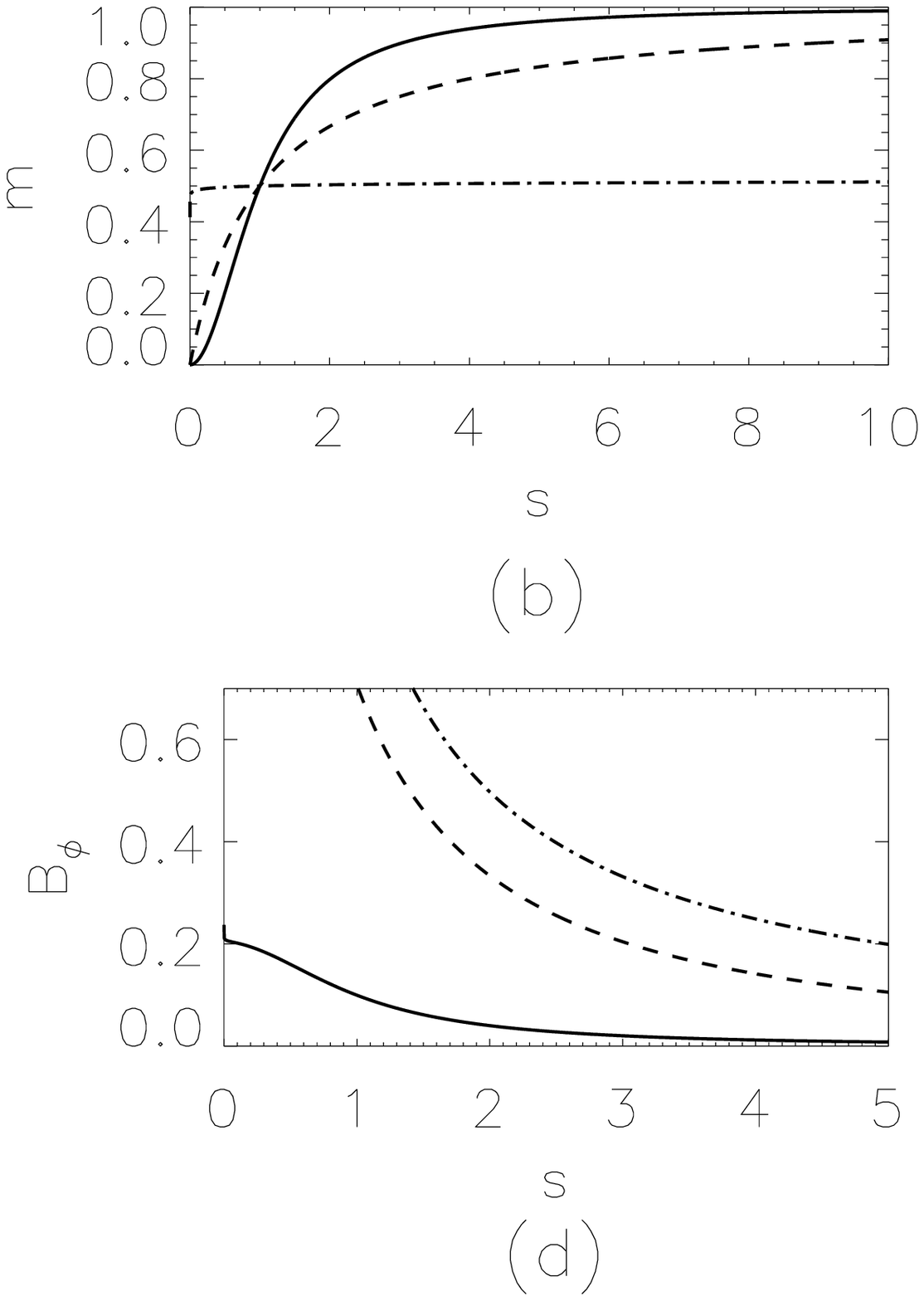}
\caption{Structure of the $\beta=\mathit{constant}$ gas.  Panel (a) shows the density structure; (b) shows the behaviour of the mass per unit length, $m$; (c) displays a log-log plot of the density; and (d) displays the toroidal magnetic field strength $\tilde{B}_{\phi}$.  The thick solid line in (c) represents the curve $\tilde{\rho}\propto s^{-2}$; the thick dashed line represents the curve $\tilde{\rho}\propto s^{-4}$ (the asymptotic solution to the non-magnetic case).\label{figbmulti}}
\end{figure}

A common description for magnetic fields in molecular clouds is a constant ratio of gas pressure to magnetic pressure, $\beta$, such that the magnetic field is $\tilde{B}_{\phi}=\left(2\beta^{-1} \tilde{\rho}\right)^{1/2}$.  We can use this in Equation (\ref{mpm}) to write

\begin{eqnarray}
\tilde{m} & = & \frac{1}{2} \pm \frac{1}{2}\sqrt{1-2s^2(1+\beta^{-1})\tilde{\rho}}
\end{eqnarray}
   We can differentiate this and use Equation (\ref{sfmass}) to solve for $\tilde{\rho}(s)$, with the result
\begin{eqnarray}
\tilde{\rho}(s) & = & \frac{\mathrm{sech}^2\left[(\ln|s|)/(1+\beta^{-1})\right]}{2s^2\left(1+\beta^{-1}\right)}\label{cbetadens}\\
\tilde{m}(s) & = & \frac{1}{2}\left\{1+\tanh\left[(\ln|s|)/(1+\beta^{-1})\right]\right\}\label{cbetamass}\\
\tilde{B}_{\phi}(s) & = & \frac{\mathrm{sech}\left[(\ln|s|)/(1+\beta^{-1})\right]}{|s|\sqrt{1+\beta}}\label{cbetamag}
\end{eqnarray}
This solution displays a number of the properties mentioned previously.  The half-mass point occurs when $|s|=1$; the value of $s_{1/2}^2(2\tilde{\rho}(s_{1/2})+\tilde{B}_{\phi}^2(s_{1/2}))=\mathrm{sech}^2(0)=1$, as expected.  This is also the sonic point in the fluid (that is, where $|\tilde{v}_r| = 1$).  The limit $\lim\limits_{|s|\rightarrow\infty}(s\tilde{B}_{\phi})=0$, so the total mass per unit length ought to be equal to the critical value $\tilde{m}_\mathrm{max}=1$; from Equation (\ref{cbetamass}) we can see that this is indeed the case.
 The solutions to Equations (\ref{cbetadens})--(\ref{cbetamag}) are plotted in Figure \ref{figbmulti}.  The behaviour of the quantity $q_\mathrm{crit}=s^2\left(2\tilde{\rho}+\tilde{B}_{\phi}^2\right)$ is plotted in Figure \ref{figcritmulti}a, and displays exactly the behaviour discussed in Section (\ref{sectioneq}).  The branches of $\tilde{m}$ are shown in Figure \ref{figcritmulti}b.

	Since most measurements of molecular clouds have difficulty resolving the interior regions of cores and filaments, we examine the behaviour of these solutions at large $s$ (equivalent to large $r$).  We can do this by expanding the $\mathrm{sech}^2$ term into exponentials.  The net result is
\begin{eqnarray}
\tilde{\rho} \propto |s|^{-2\frac{2\beta+1}{\beta+1}}\label{cbetalarges}
\end{eqnarray}
for $|s|\gg 1$.  Equation (\ref{cbetalarges}) shows that for $\beta\gg 1$ (corresponding to weak magnetic fields), $\tilde{\rho}\propto s^{-4}$.  For $\beta\ll 1$ (strong magnetic fields), $\tilde{\rho} \propto s^{-2}$.  Note that in the limit $s\rightarrow \infty$ (equivalent to $t\rightarrow0$), $\tilde{\rho}\rightarrow 0$ for all $r\ne 0$.  This simply shows that there cannot be an accretion solution for $t>0$---there is no material left to be accreted.

	We can similarly write an expression for the density behaviour as $s\rightarrow0$:
\begin{eqnarray}
\tilde{\rho} \propto |s|^{-\frac{2}{1+\beta}}
\end{eqnarray}
For a finite value of $\beta$, the density will be singular at the axis.  This is in contrast to the result of \citet{stodolkiewicz63}, who used an arbitrary boundary condition of $\rho(r=0)=\mathit{constant}$.  We believe our result to be physically more self-consistent, as it satisfies the physical constraints of zero mass per unit length and toroidal flux at the origin (the solution of \citet{stodolkiewicz63} cannot simultaneously satisfy the hydrodynamic equations and the restriction of $m(0)=0$, for a purely toroidal field).  Our solution shows that the density in a $\beta=\mathrm{constant}$ model will be singular on the axis.  A singularity in the spherical solutions can be rationalized as an approximation to the effects of a protostar in the center of the cloud. In the cylindrical case, even though collapse as a singular filament occurs, this structure itself will undergo fragmentation in a finite amount of time, as is shown in Section \ref{sectionfrag}.  Our treatment of the $\beta=\mathrm{constant}$ case shows that it may be of limited astrophysical interest.

	The toroidal field contributes both a magnetic pressure force and a magnetic tension force to the momentum equations.  It is instructive to compare the magnitudes of these two forces.  The magnetic pressure is $(1/2)\;\mathrm{d}(\tilde{B}_{\phi}^2)/\mathrm{d}s$, and the magnetic tension force is $-\tilde{B}_{\phi}^2/s$.  The ratio is thus
\begin{eqnarray}
R & = & \frac{1}{2}\frac{\mathbf{\nabla}\mathbf{\tilde{B}}^2}{(\mathbf{\tilde{B}\cdot\nabla})\mathbf{\tilde{B}}} = \frac{1}{2}\frac{\partial \ln (\tilde{B}_{\phi}^2)}{\partial \ln |s|}
\end{eqnarray}
For our $\beta=\mathit{constant}$ fluid, $R=(1/2)\;\mathrm{d}(\ln\tilde{\rho})/\mathrm{d}(\ln |s|) = 2\left(2\tilde{m}+\beta^{-1}\right)$.  The magnetic pressure force will dominate for $|R| > 1$, which corresponds to $\tilde{m}>1/4-(1/2)\beta^{-1}$; note that this will always be the case for a fluid with $\beta < 2$, as then the right-hand side will be negative.  Strongly magnetized filaments are therefore supported by the pressure forces associated with their toroidal fields.

\subsection{Case 1B: Constant Flux-to-Mass Ratio}

\begin{figure}
\plottwo{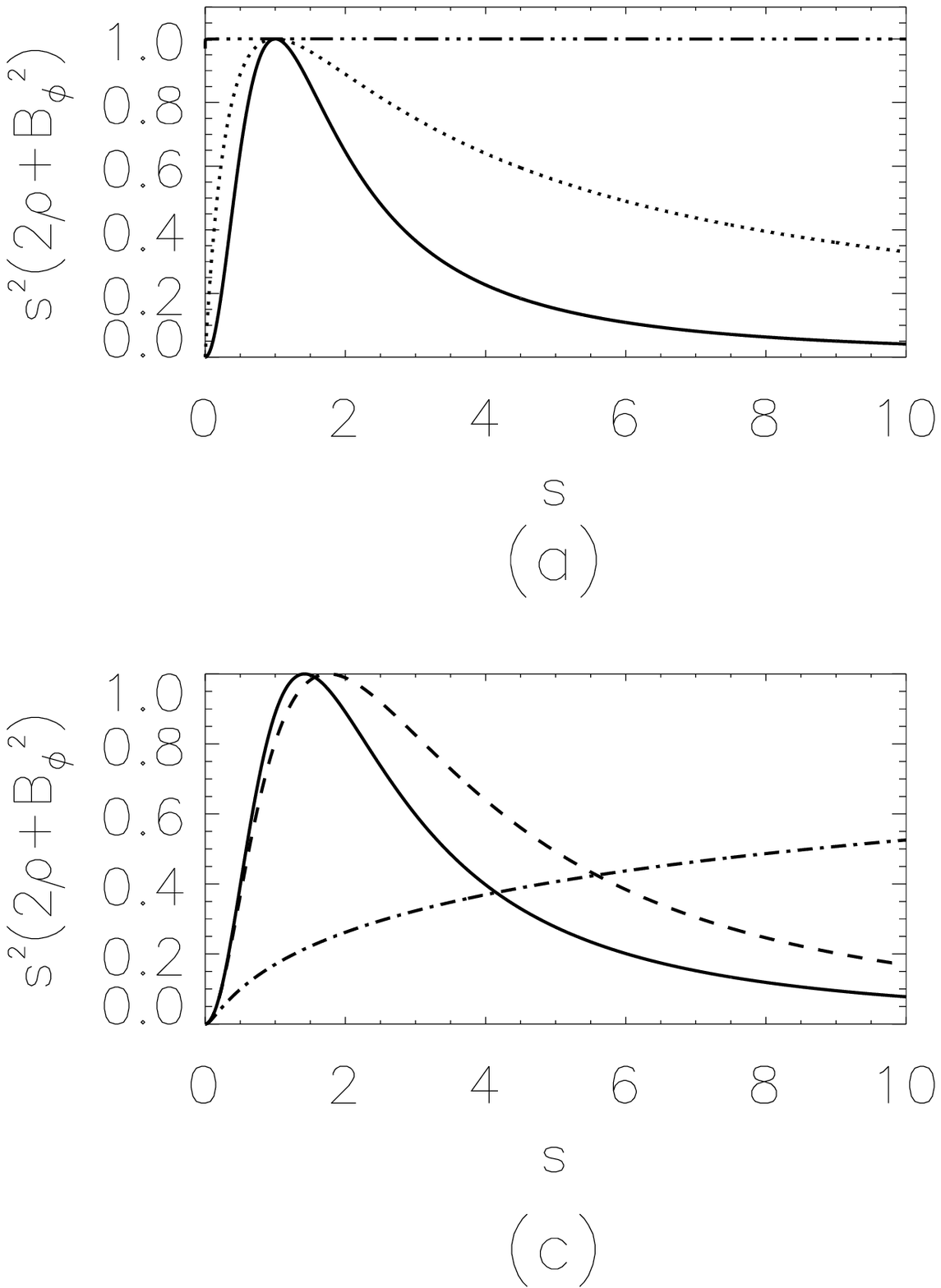}{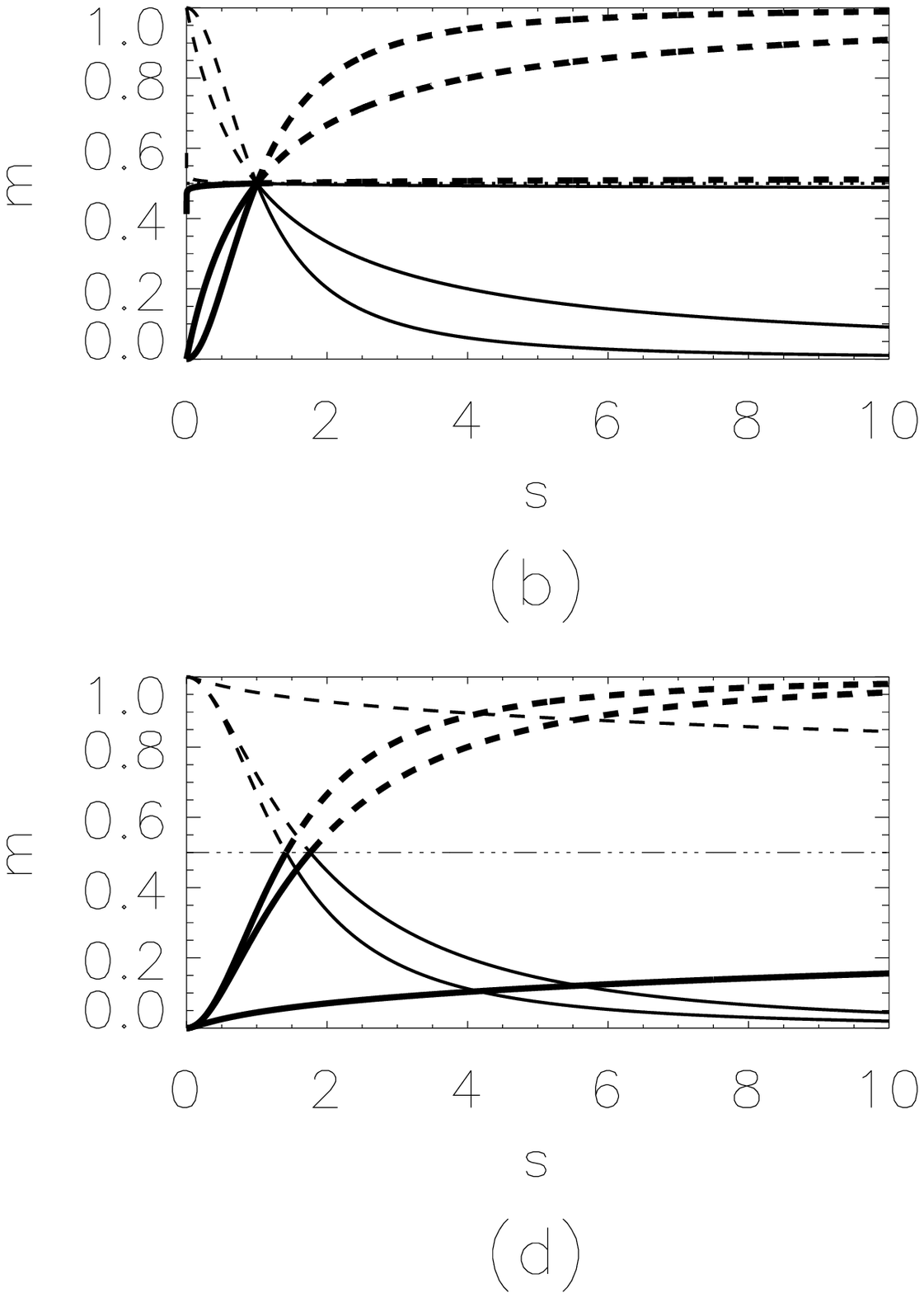}
\caption{Behaviour of the critical quantity $q_\mathrm{crit}$ and the mass per unit length $\tilde{m}$.  The upper panels show the behaviour of the $\beta=\mathit{constant}$ gas; the lower panels are for the $\Gamma_{\phi}=\mathit{constant}$ case.  The panels on the left display the behaviour of the critical quantity $s^2(2\tilde{\rho}+\tilde{B}_{\phi}^2)$; those on the right show the upper branches (dotted lines) and lower branches (solid lines) of the mass per unit length.  The heavy lines mark the allowable branches.  The different lines in (a) describe the same models as the lines in Figure \ref{figbmulti}; the different lines in (c) describe the same models as the lines in Figure \ref{figgmulti}.\label{figcritmulti}}
\end{figure}

\begin{figure}
\plottwo{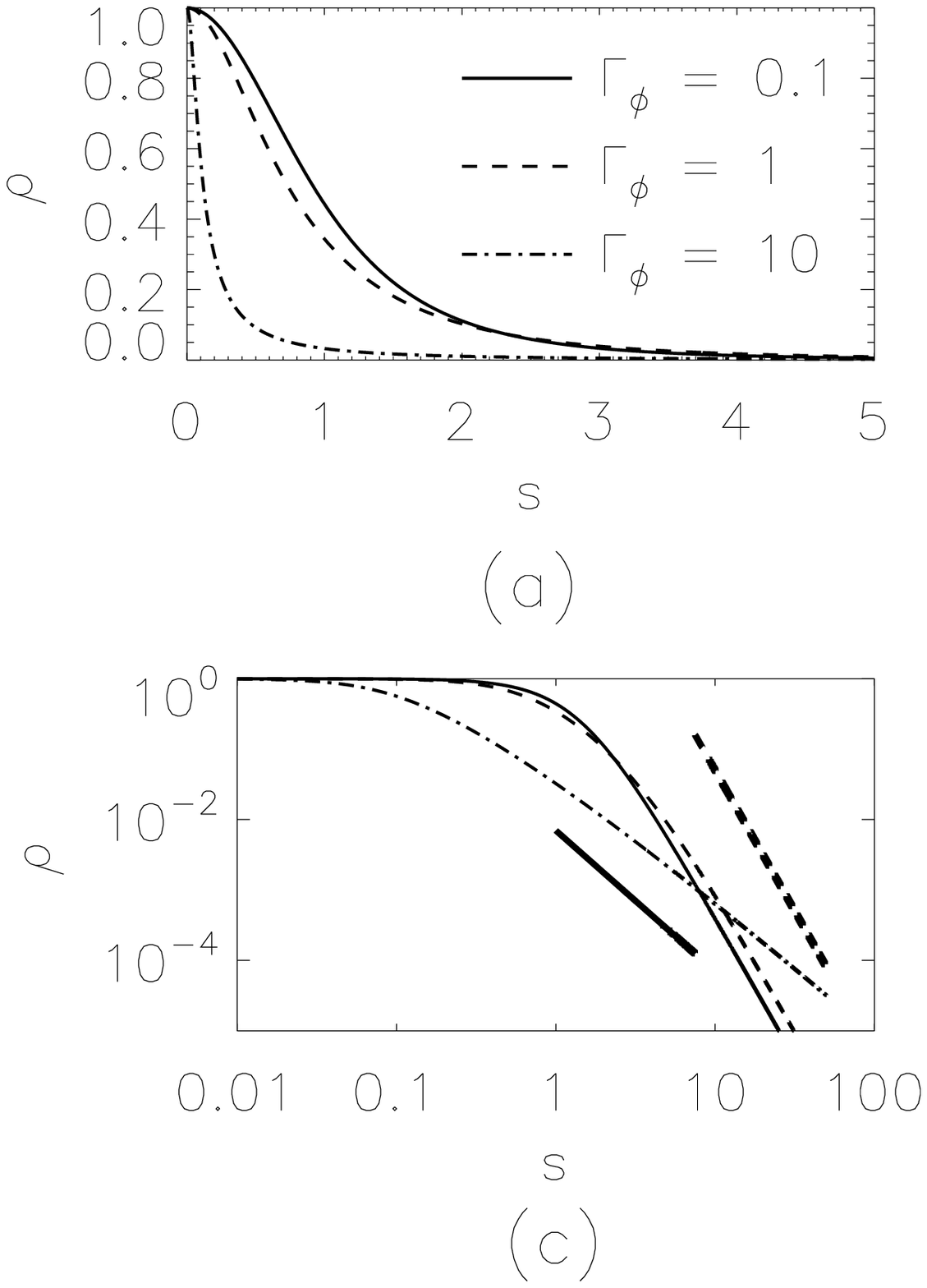}{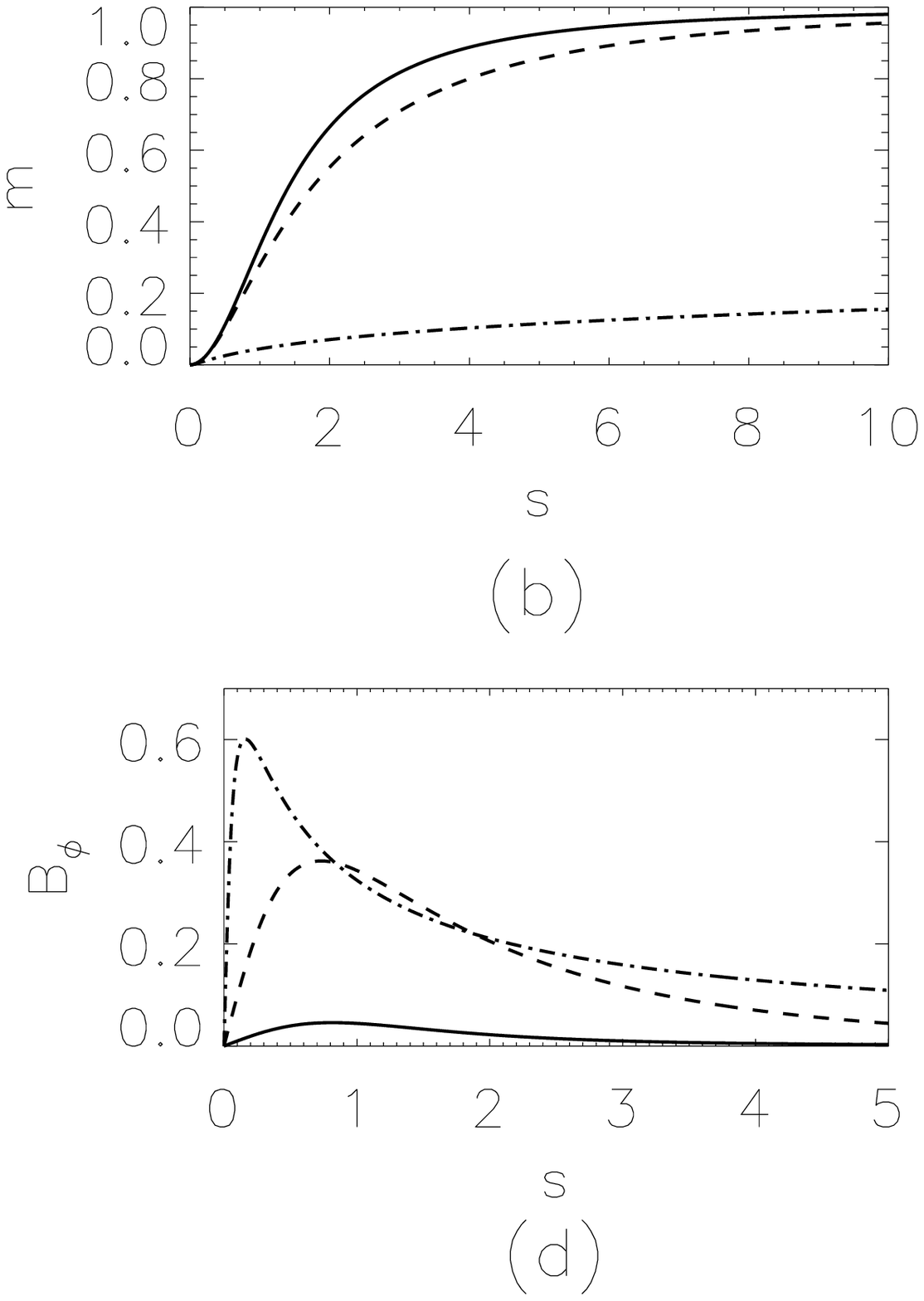}
\caption{Structure of the $\Gamma_{\phi}=\mathit{constant}$ gas.  Panel (a) shows the density structure; (b) shows the behaviour of the mass per unit length $m$; (c) displays a log-log plot of the density; and (d) displays the toroidal magnetic field strength $\tilde{B}_{\phi}$.
The thick solid line in (c) represents the curve $\tilde{\rho}\propto s^{-2}$; the thick dashed line represents  the curve $\tilde{\rho}\propto s^{-4}$ (the asymptotic solution to the non-magnetic case).\label{figgmulti}}
\end{figure}

  Another form for the magnetic field configuration of a filament that has been proposed \citep{fiege00a} is one wherein the magnetic flux through a volume is proportional to the mass in the volume, such that the flux-to-mass ratio $\Gamma_{\phi}$ is constant everywhere ($\tilde{B}_{\phi} = \Gamma_{\phi} |s| \rho$). With this assumption, we can numerically integrate Equation (\ref{sfmom}) using a Runge-Kutta integrator to obtain a set of solutions.  We plot these solutions for several values of $\Gamma_\phi$ in Figure \ref{figgmulti}.  As for the solutions for the $\beta=\mathit{constant}$ plasma, the solutions for $\Gamma_\phi=\mathit{constant}$ exhibit the properties discussed in Section (\ref{sectioneq}).  However, we also note some differences in the behaviour of these two different systems.  We plot the critical quantity $q_\mathrm{crit}$ for $\Gamma_\phi=\mathit{constant}$ in Figure \ref{figcritmulti}c, which can be directly compared to the $q_\mathrm{crit}$ of the $\beta=\mathit{constant}$ gas, plotted in the panel above it.  While the half-mass point for the $\beta=\mathit{constant}$ case did not depend on the value of the parameter $\beta$, for the $\Gamma_\phi=\mathit{constant}$ solution the half-mass point does depend on the value of $\Gamma_\phi$.  The density and magnetic field for the $\Gamma_\phi=\mathit{constant}$ solution are also non-singular, as can be seen in Figure \ref{figgmulti}.

	At low values of magnetization (i.e. low $\Gamma_{\phi}$), the density distribution is very similar to that predicted by \citet{miyama87} for the self-similar collapse of an unmagnetized filament (which in turn has a the same shape as the unmagnetized hydrostatic filament).  As the magnetic flux increases, the density distribution at large $|s|$ tends towards $\tilde{\rho} \propto s^{-2}$ rather than the steeper $s^{-4}$ for weak magnetic fields.  This can be shown explicitly by isolating $d\tilde{\rho}/ds$ in Equation (\ref{sfmom}):
\begin{eqnarray}
\frac{d\tilde{\rho}}{ds} & = & -\left(\frac{2\tilde{\rho}}{s}\right)\frac{2\tilde{m}+\Gamma_{\phi}^2 s^2\tilde{\rho}}{1+\Gamma_{\phi}^2 s^2\tilde{\rho}}
\end{eqnarray}
For $\Gamma_{\phi}\ll 1$, we see that $d\ln(\tilde{\rho})/d\ln|s| = -4\tilde{m}$; since $m\simeq 1$ as $|s|\rightarrow\infty$, $\tilde{\rho}\propto s^{-4}$ in the case of low magnetization.  For $\Gamma_{\phi}\gg 1$, we see that $d\ln(\tilde{\rho})/d\ln|s| = -2$, implying that $\tilde{\rho}\propto s^{-2}$, as discussed above.  These results and those in Section \ref{subsectionconstantbeta} show that the asymptotic behaviour at large radii for very weak or very strong magnetic fields is independent of the details of how these filaments are magnetized.

As for the constant-$\beta$ solution, as $s\rightarrow\infty$, $\tilde{\rho}\rightarrow0$.  Thus, for solutions with a constant $\Gamma_{\phi}$, there cannot be an accretion solution for $t>0$, either.

\subsection{Fragmentation}\label{sectionfrag}

   \citet{stodolkiewicz63} and \citet{fiege00b} derived criteria for the fragmentation of a magnetized isothermal cylinder, for each of the forms of the magnetization that we have used.  Since our collapse solution moves towards a line singularity at t=0, which is unphysical, there exists the possibility that the cylinder will become unstable to fragmentation instabilities as the density and magnetic field strengths change with time.

  \citet{fiege00b} found that for a filament with constant toroidal flux-to-mass ratio, the fragmentation timescale is
\begin{eqnarray}
\tau_{frag} = 1.8 \left(-\frac{\omega_{max}^2}{0.01}\right)^{-1/2}\left(\frac{n_c}{10^4 \mathrm{cm}^{-3}}\right) \mathrm{Myr}
\end{eqnarray}
where $w_{max}^2$ is a frequency of fragmentation which depends on the value of $\Gamma_{\phi}$.  \citet{fiege00b} found that $-\omega_{max}^2 \le 0.0581$.

  As the radial density structure is the same in both the self-similar case and the hydrostatic case, we can compare the fragmentation timescale to our collapse timescale, to see which process should dominate.  The collapse timescale is
\begin{eqnarray}
\tau_{col} = \frac{1}{\sqrt{\pi G \rho_c}} \sim 0.38\left(\frac{n_c}{10^4 \mathrm{cm}^{-3}}\right)^{-1/2} \mathrm{Myr}
\end{eqnarray}
and so the ratio of the fragmentation to collapse timescales is
\begin{eqnarray}
\frac{\tau_{frag}}{\tau_{col}} = 4.75\left(-\frac{\omega_{max}^2}{0.01}\right)^{-1/2}
\end{eqnarray}
This ratio has a minimum at $\Gamma_{\phi}=0$ of $\tau_{frag}/\tau_{col} = 1.97$, which increases as $\Gamma_{\phi}$ increases.  Thus, the toroidal magnetic field acts to prevent the cloud from fragmenting.  In the case of low magnetization, the collapse and fragmentation timescales are approximately equal, suggesting that the cloud may begin to collapse along the filament axis, in addition to its radial collapse.  

  We can also find the ratio of the critical wavelength for fragmentation to the radial collapse scale.  \citet{fiege00b} found that the fragmentation wavelength for a constant toroidal flux-to-mass ratio is
\begin{eqnarray}
\lambda_{frag}=2.8\left(\frac{\sigma_c}{0.5 \mathrm{km s}^-1}\right)\left(\frac{n_c}{10^4 \mathrm{cm}^{-3}}\right)^{-1/2}\left(\frac{k_{max}}{0.2}\right)^{-1} \mathrm{pc}
\end{eqnarray}
where $k_{max}$ is a dimensionless wavenumber corresponding to the maximum instability, and has a maximum value of 0.462 at $\Gamma_{\phi}=0$.  An estimate of the radial length scale can be set by setting $|s|=1$:
\begin{eqnarray}
\lambda_{rad} & = & c_s s t \nonumber \\
\lambda_{rad} & = & c_s \frac{1}{\sqrt{\pi G \rho_c}} \nonumber \\
\lambda_{rad} & = & 0.194\left(\frac{\sigma_c}{0.5 \mathrm{km s}^-1}\right)\left(\frac{n_c}{10^4 \mathrm{cm}^{-3}}\right)^{-1/2} \mathrm{pc}\label{lambdarad}
\end{eqnarray}
From this, we get a ratio of fragmentation to radial wavelengths of
\begin{eqnarray}
\frac{\lambda_{frag}}{\lambda_{rad}} = 14.43 \left(\frac{k_{max}}{0.2}\right)^{-1}
\end{eqnarray}
At $\Gamma_{\phi}=0$, this has its minimum value of 6.25 (3.41 if we use the critical wavelength rather than the maximum wavelength).  Again, this suggests that toroidal fields support the cylinder against collapse.

  We can do a similar analysis for the $\beta=\mathit{constant}$ case.  The fragmentation wavelength found by \citet{stodolkiewicz63} for a $\beta=\mathit{constant}$ fluid is
\begin{eqnarray}
\lambda_{frag}=1.08\left(\frac{\sigma_c}{0.5 \mathrm{km s}^-1}\right)\left(\frac{n_c}{10^4 \mathrm{cm}^{-3}}\right)^{-1/2}\sqrt{1+\frac{1}{\beta}} \quad \mathrm{pc}
\end{eqnarray}
while the radial wavelength is still given by Equation (\ref{lambdarad}), for a ratio of
\begin{eqnarray}
\frac{\lambda_{frag}}{\lambda_{rad}} = 5.563\sqrt{1+\frac{1}{\beta}}
\end{eqnarray}
Again, this has a minimum value when the magnetization is zero ($\beta=\infty$); the presence of a toroidal field serves to stabilize the cylinder against fragmentation.

	In addition to the axisymmetric instabilities discussed above, there exists the possibility that non-axisymmetric instabilities (e.g. kink modes) could develop.  The presence of such instabilities would destroy the axisymmetric nature of our models, a problem that cannot be addressed with a self-similar formalism.  We can, however, estimate the likelihood of such modes by comparing their growth rates with that of the axisymmetric modes.  \citet{nakamura93} has investigated the kink mode instabilities of a hydrostatic molecular cloud with helical magnetic fields described by a constant $\beta$.  They found that the axisymmetric instabilities, such as those described above, have faster growth rates than all non-axisymmetric modes.  Thus, axisymmetric modes will dominate the fragmentation, and non-axisymmetric modes are not an important effect in the evolution of these filaments on the collapse time scales that we are considering.

\section{DISCUSSION AND CONCLUSIONS}

  We have determined the structure of a collapsing self-similar magnetized isothermal cylinder for the two most common magnetic field configurations found in the literature.  One of the major conclusions of our study is that the cylindrical radial structure for the gas density is indistinguishable from the hydrostatic structure with the same magnetic field configuration.  Consequently, one would need to examine the profiles of molecular lines for infall signatures in order to differentiate collapsing and equilibrium structures.  Because the high-density regions that emit most of the thermal radiation will have low velocities, it may be more difficult to detect a characteristic line profile in comparison with the situation for accreting spherical cores that can be identified by an inverse P-Cygni profile.

	The self-similar density profile at large radii from the axis follows a power law.  For a cylinder with a weak magnetic field, we find that at any particular time during the collapse, $\rho \propto r^{-4}$.  By way of contrast, for a cylinder with a strong magnetic field, we find that $\rho \propto r^{-2}$.  These results are independent of the model for the magnetic field strength that we use ($\beta = \mathit{constant}$ or $\Gamma_{\phi}=\mathit{constant}$).

	Our solutions require that there be no motions along the axis, and that the magnetic field is purely toroidal.  In a fully dynamic, three-dimensional collapse, we would not necessarily expect this to be the case.  However, if the axial motions and poloidal fields are sufficiently weak, we could reasonably expect our solutions to be a good approximation.  Furthermore, the static models of \citet{fiege00a} indicate that for models consisting of both poloidal and toroidal fields, the toroidal magnetic field strongly dominates in the outer regions of a filament, suggesting that our solutions may be applicable to the envelopes of filaments threaded with helical fields.

  The velocity behaviour of our solutions bears a close resemblence to the solution found by \citet{penston69} for the self-similar collapse of an isothermal sphere, in that the infall velocity $v_r \rightarrow 0$ as $r \rightarrow 0$.  However, due to the simple nature of the continuity equation, we only find the collapse solution; there is no analogue to the \citet{shu77} accretion--expansion wave solution that follows the formation of a singular core.  It is also the continuity equation which requires $v\propto r$ at all radii, rather than having an asymptotic infall velocity as found by \citet{penston69}.  Clearly, arbitrarily high velocities at large radii are not physical; there needs to be a mechanism to truncate the models at finite $r$.  Pressure truncation, which we have not included in this analysis, could satisfy this condition.

  We find, in agreement with \citet{stodolkiewicz63} and \citet{fiege00b}, that the presence of a toroidal field stabilizes the cylinder against fragmentation.  Even in the absence of a field, however, collapse motions should dominate over fragmentation.  We thus have shown that self-similar filaments should remain relatively unfragmented as they collapse.  This mechanism would allow filamentary structures to survive over long periods of time while retaining their filamentary character.  If this were not true, we would not expect to see many filaments because they would fragment into cores too quickly.

	Although large scale velocity fields have been observed to be associated with individual cloud cores (e.g. \citeauthor{mardones97} \citeyear{mardones97}; \citeauthor{lee99} \citeyear{lee99}), we emphasize that the flows predicted here are on a yet larger scale associated with filament formation.  Simulations of hydromagnetic turbulence show that filaments can form from shocks \citep{ostriker99,ostriker01}.  We plan to investigate the possible link between our analytic models and time-dependent simulations of magnetized clouds.

	We thank an anonymous referee for useful comments on the manuscript.  D.A.T. is supported by an Ontario Graduate Scholarship.  R.E.P. is supported by the Natural Sciences and Engineering Research Council of Canada.

\appendix
\section{THE STRUCTURE OF THE SELF-SIMILAR AND HYDROSTATIC EQUATIONS}\label{appendixa}

We can show explicitly that the structure of a filament that is collapsing self-similarly is identical to that of a hydrostatic filament with the same constitutive equations for the magnetic field.  In our formalism, we replace the gravitational potential with the mass per unit length (Equation \ref{rtmass})
\begin{eqnarray}
r\frac{\partial \Phi}{\partial r} = 2 G m
\end{eqnarray}
We write out Equations (\ref{continuity})-(\ref{poisson}) in component form, and use the fact that we are considering only the axisymmetric and axially symmetric case (i.e. functions depend only on the radius r, and the time t).
\begin{eqnarray}
\frac{\partial m}{\partial t} + v_r \frac{\partial m}{\partial r} & = & 0\\
\frac{\partial B_r}{\partial t} & = & 0\\
\frac{\partial B_{\phi}}{\partial t} & = & -\frac{\partial}{\partial r}\left(v_r B_{\phi}-v_{\phi} B_r\right)\\
\frac{\partial B_z}{\partial t} & = & \frac{1}{r}\frac{\partial}{\partial r}\left[r\left(v_z B_r - v_r B_z\right)\right]\\
\rho \frac{\partial v_r}{\partial t} + \rho v_r \frac{\partial v_r}{\partial r} - \rho \frac{v_{\phi}^2}{r} & = &  \frac{1}{\mu}\left[B_r\frac{\partial B_r}{\partial r} - \frac{B_{\phi}^2}{r}-\frac{1}{2}\frac{\partial}{\partial r}\left(B_r^2+B_{\phi}^2+B_z^2\right)\right] \nonumber\\
&  & - c_s^2 \frac{\partial \rho}{\partial r}-\frac{2 G m\rho}{r} \\
\rho\frac{\partial v_{\phi}}{\partial t} + \rho\left(v_r\frac{\partial v_{\phi}}{\partial r}+\frac{v_{\phi} v_r}{r}\right) & = & \frac{1}{\mu}\left[B_r \frac{\partial B_{\phi}}{\partial r} + \frac{B_{\phi} B_r}{r}\right]\\
\rho \frac{\partial v_z}{\partial t}+\rho v_r \frac{\partial v_z}{\partial r} & = & \frac{B_r}{\mu}\frac{\partial B_z}{\partial r}\label{rtmomz}
\end{eqnarray}
If we make the substitutions given in Equations (\ref{rhortos})-(\ref{brtos}), we obtain the self-similar structure equations (\ref{scont})-(\ref{smomphi}).  If we instead consider the equilibrium solutions (so that $\mathbf{v}=\mathbf{0}$, ${\partial/\partial t=0}$), we get
\begin{eqnarray}
0 & = & \frac{1}{\mu}\left[\frac{B_{\phi}}{r}\frac{d}{d r}\left(r B_{\phi}\right)+B_z \frac{d B_z}{d r}\right] + c_s^2 \frac{d \rho}{d r} + \frac{2 G m \rho}{r} \\
0 & = & B_r \frac{d}{d r}\left(r B_{\phi}\right) \\
0 & = & B_r \frac{d B_z}{d r} \\
\frac{dm}{dr} & = & 2\pi r\rho
\end{eqnarray}
From the condition that $\mathbf{\nabla \cdot B} = 0$, we require $B_r = 0$, for the same reasons we had in Section \ref{sectioneq} for Equation (\ref{sindr}).  We now convert to dimensionless quantities, using (in analogy with Equations \ref{rhortos}-\ref{brtos})
\begin{eqnarray}
\rho & = & \frac{1}{\pi G}\bar{\rho} \\
m & = & \frac{2 c_s^2}{G}\bar{m} \\
B_{\phi} & = & c_s \sqrt{\frac{\mu}{\pi G}} \bar{B}_{\phi} \\
B_z & = & c_s \sqrt{\frac{\mu}{\pi G}} \bar{B}_z \\
r & = c_s x
\end{eqnarray}
This results in
\begin{eqnarray}
\frac{d\bar{m}}{dx} & = & x\bar{\rho} \label{xfmass}\\
0 & = & \frac{d\bar{\rho}}{dx} + \frac{4\bar{m}\bar{\rho}}{x} + \frac{\bar{B}_{\phi}}{x}\frac{d}{dx}\left(x\bar{B}_{\phi}\right) + \frac{d \bar{B}_z}{dx} \label{xfmom}
\end{eqnarray}
Comparing Equations (\ref{xfmass}) and (\ref{xfmom}) with Equations (\ref{sfmass}) and (\ref{sfmom}), we see that in the cylindrically symmetric geometry, the self-similar collapse equations and hydrostatic equilibrium equations will always give the same structure for the density (with the exception that a $B_z$ field in the hydrostatic case is allowed), given a constitutive relation between the toroidal magnetic field and the density.


\bibliographystyle{apj}
\bibliography{ref}

\begin{thebibliography}{20}
\expandafter\ifx\csname natexlab\endcsname\relax\def\natexlab#1{#1}\fi

\bibitem[{{Falgarone} {et~al.}(2001){Falgarone}, {Pety}, \&
  {Phillips}}]{falgarone01}
{Falgarone}, E., {Pety}, J., \& {Phillips}, T.~G. 2001, \apj, 555, 178

\bibitem[{{Fiege} \& {Pudritz}(2000{\natexlab{a}})}]{fiege00a}
{Fiege}, J.~D. \& {Pudritz}, R.~E. 2000{\natexlab{a}}, MNRAS, 311, 85

\bibitem[{{Fiege} \& {Pudritz}(2000{\natexlab{b}})}]{fiege00b}
---. 2000{\natexlab{b}}, MNRAS, 311, 105

\bibitem[{{Foster} \& {Chevalier}(1993)}]{foster93}
{Foster}, P.~N. \& {Chevalier}, R.~A. 1993, ApJ, 416, 303+

\bibitem[{{Heiles} {et~al.}(1993){Heiles}, {Goodman}, {McKee}, \&
  {Zweibel}}]{heiles93}
{Heiles}, C., {Goodman}, A.~A., {McKee}, C.~F., \& {Zweibel}, E.~G. 1993, in
  Protostars and Planets III, 279--326

\bibitem[{{Hennebelle}(2002)}]{hennebelle02}
{Hennebelle}, P. 2002, {preprint, astro-ph/0210422}

\bibitem[{{Klessen} \& {Burkert}(2000)}]{klessen00a}
{Klessen}, R.~S. \& {Burkert}, A. 2000, ApJS, 128, 287

\bibitem[{{Larson}(1969)}]{larson69}
{Larson}, R.~B. 1969, MNRAS, 145, 271+

\bibitem[{{Lee} {et~al.}(1999){Lee}, {Myers}, \& {Tafalla}}]{lee99}
{Lee}, C.~W., {Myers}, P.~C., \& {Tafalla}, M. 1999, \apj, 526, 788

\bibitem[{{Mardones} {et~al.}(1997){Mardones}, {Myers}, {Tafalla}, {Wilner},
  {Bachiller}, \& {Garay}}]{mardones97}
{Mardones}, D., {Myers}, P.~C., {Tafalla}, M., {Wilner}, D.~J., {Bachiller},
  R., \& {Garay}, G. 1997, \apj, 489, 719

\bibitem[{{Matthews} \& {Wilson}(2000)}]{matthews00}
{Matthews}, B.~C. \& {Wilson}, C.~D. 2000, \apj, 531, 868

\bibitem[{{Miyama} {et~al.}(1987){Miyama}, {Narita}, \& {Hayashi}}]{miyama87}
{Miyama}, S.~M., {Narita}, S., \& {Hayashi}, C. 1987, Prog. Theor. Phys., 78,
  1051+

\bibitem[{{Nakamura} {et~al.}(1993){Nakamura}, {Hanawa}, \&
  {Nakano}}]{nakamura93}
{Nakamura}, F., {Hanawa}, T., \& {Nakano}, T. 1993, \pasj, 45, 551

\bibitem[{{Ostriker} {et~al.}(1999){Ostriker}, {Gammie}, \&
  {Stone}}]{ostriker99}
{Ostriker}, E.~C., {Gammie}, C.~F., \& {Stone}, J.~M. 1999, \apj, 513, 259

\bibitem[{{Ostriker} {et~al.}(2001){Ostriker}, {Stone}, \&
  {Gammie}}]{ostriker01}
{Ostriker}, E.~C., {Stone}, J.~M., \& {Gammie}, C.~F. 2001, ApJ, 546, 980

\bibitem[{{Ostriker}(1964)}]{ostriker64}
{Ostriker}, J. 1964, ApJ, 140, 1056+

\bibitem[{{Penston}(1969)}]{penston69}
{Penston}, M.~V. 1969, MNRAS, 144, 425+

\bibitem[{{Porter} {et~al.}(1994){Porter}, {Pouquet}, \& {Woodward}}]{porter94}
{Porter}, D.~H., {Pouquet}, A., \& {Woodward}, P.~R. 1994, Phys. Fluids, 6,
  2133

\bibitem[{{Shu}(1977)}]{shu77}
{Shu}, F.~H. 1977, ApJ, 214, 488

\bibitem[{{Stod\'{o}{\l}kiewicz}(1963)}]{stodolkiewicz63}
{Stod\'{o}{\l}kiewicz}, J.~S. 1963, Acta Astronomica, 13, 30

\end{thebibliography}
\end{document}